\documentclass[fleqn]{article}

\usepackage{amsmath,amssymb}
\usepackage{graphicx}
\usepackage[top=0.75in, bottom=0.75in, left=0.75in, right=0.75in]{geometry}
\usepackage[small]{caption}
\usepackage[noblocks]{authblk}
\usepackage{hyperref}
\usepackage{ulem}
\usepackage{color}

\begin{document}

\title{{\sf  QCD Sum-Rule Interpretation of  X(3872) with  $J^{PC}=1^{++}$ Mixtures of  Hybrid Charmonium and $\bar D D^*$ Molecular Currents }}

\author[2]{Wei Chen}
\author[1]{Hong-ying\ Jin}
\author[2]{R.T.\ Kleiv}
\author[2]{T.G.\ Steele}
\author[1]{Meng Wang}
\author[1]{Qing Xu}

\affil[1]{Zhejiang Institute of Modern Physics, Zhejiang University, Zhejiang Province, P. R. China}
\affil[2]{Department of Physics and Engineering Physics, University of Saskatchewan, Saskatoon, Saskatchewan, S7N 5E2, Canada}

\maketitle
\begin{abstract}
QCD sum-rules are employed to determine whether the X(3872) can be described as a mixed state that couples to
$J^{PC}=1^{++}$ charmonium hybrid and $\bar D D^*$ molecular currents.  After calculating the mixed correlator of hybrid and molecular currents, we formulate the sum-rule in terms of a mixing parameter that interpolates between the pure molecular and hybrid
scenarios.   As the mixing parameter is increased from the pure molecular case, the predicted mass increases until it reaches a maximum value in good agreement with the X(3872) and the resulting sum-rule analysis appears more robust than the pure molecular  case.  
\end{abstract}

\section{Introduction}
\label{theIntroduction}

The X(3872) state was first detected in August 2003 by Belle \cite{Choi:2003ue}
in $B^{\pm}\rightarrow\pi^{+}\pi^{-}J/\psi K^{\pm}$, and its existence
was quickly confirmed by the CDF \cite{Acosta:2003zx}, D0 \cite{Abazov:2004kp}
and BaBar \cite{Aubert:2004fc} collaborations. The observation of
the X(3872) and other new states with 
masses in the range
 3.8--4.7~GeV
has led to a resurgence of interest in exotic meson spectroscopy \cite{Swanson:2006st}.
Recently the LHCb collaboration has measured the X(3872)  mass as $m_{X(3872)}=3871.95\pm0.48(\text{stat})\pm0.12(\text{syst})\,{\rm MeV}$ \cite{Aaij:2011sn}.
The PDG  summary table has listed the X(3872)
with an average mass $m{}_{X(3872)}=3871.68\pm0.17 \,{\rm MeV}$ and a narrow
width $\Gamma_{X(3872)}<1.2\,{\rm MeV}$ \cite{Beringer:1900zz}.
The radiative decay $X(3872)\rightarrow\gamma J/\psi$ \cite{Abe:2005ix,Choi:2011fc,Aubert:2006aj,Aubert:2008ae}
implies the C-parity of the X(3872) is even.
The LHCb collaboration has recently provided evidence to preclude $J^{PC}=2^{-+}$ from
 the $1^{++}$ and $2^{-+}$ assignments \cite{Aaij:2013zoa}
  (Ref.~\cite{Hanhart:2011tn} is  another recent
work  supporting $J^{PC}=1^{++}$).

 With strong evidence supporting the $1^{++}$ quantum numbers,
  the X(3872) does not  appear to be a  pure conventional
charmonium state because the  mass of  $1^{++}$ $\chi_{c1}(2P)$ from the quark model calculation is too
 large.
 The $1^{++}$ assignment of  the
X(3872), due to its exotic  interpretation,  has attracted a lot of attention.
 To date,  many  structures for the X(3872) have been proposed,
such as a tetraquark state \cite{Maiani:2004vq,Ebert:2005nc,Matheus:2006xi,Terasaki:2007uv,Dubnicka:2010kz},
$\overline{c}cg$ hybrid state \cite{Li:2004sta}, charmonium \cite{Barnes:2003vb,Suzuki:2005ha}
and a $D^{0}\overline{D}^{*0}$ molecule state \cite{Close:2003sg,Voloshin:2003nt,Swanson:2003tb,Tornqvist:2004qy,AlFiky:2005jd,Thomas:2008ja,Liu:2008tn,Lee:2009hy,Gamermann:2009uq}  (see Refs.~\cite{Swanson:2006st,Brambilla:2010cs} for  comprehensive reviews).
The $\bar D D^*$ molecule picture  was  developed before the experimental observation, so it  attracted  considerable  attention,  but the  molecular scenario   is hard to
 reconcile with the large radiative $J/\psi \pi\pi$ decay rate  because the loosely bound $\bar D D^*$  is unlikely to annihilate into $J/\psi \pi\pi$.

The tetraquark  or hybrid  scenarios may explain  the decay rate puzzle.  But the expected mass of the pure axial-vector charmonium hybrid is about 5~GeV \cite{Harnett:2012gs}, well above the X(3872).
 Furthermore, the diquark-antidiquark (tetraquark) scenario should have partners of X(3872),  with the charged partner possibly identified as the $Z_c(3895)$ seen by BES and Belle \cite{Ablikim:2013mio,Liu:2013dau} (see also Ref.~\cite{Xiao:2013iha}).
More recently, a mixing scenario
 has been proposed \cite{Coito:2010if,Coito:2012vf,Coito:2010cq,Coito:2012ka,Takizawa:2012hy}, where it  is  suggested that the $c\bar c$  $1^{++}$ $\chi_{c1}(2P)$ strongly couples to $\bar D D^*$ so that its mass is dynamically shifted about 100 MeV downwards from the bare
$\chi_{c1}(2P)$ state. Thus this scenario may both solve both the mass and decay width puzzles.

QCD sum-rules have been used to study a number of $1^{++}$ scenarios for the  X(3872)
\cite{Matheus:2006xi,Chen:2012pe,Chen:2010ze,Nielsen:2008,Matheus:2009vq}.  It is difficult to assess whether the tetraquark or molecular currents  provide an adequate description of X(3872).  For example, Refs.~\cite{Chen:2012pe,Chen:2010ze} find a tetraquark mass of about $4.2\,{\rm GeV}$ and a molecular mass of about  $4.1\,{\rm GeV}$, while
Refs.~\cite{Matheus:2006xi,Nielsen:2008} find nearly degenerate molecular and tetraquark masses of about   $3.9\,{\rm GeV}$.  However, in these analyses it is not possible to find a sum-rule window of validity unless a 50\% continuum contribution is permitted.
Mass predictions for $1^{++}$ states  seem to be  insensitive to the choice of tetraquark or molecular currents \cite{Narison:2010pd},  but lead to different consequences for  branching ratios \cite{Karliner:2010sz}.  However,  the tetraquark and molecular currents are related through Fierz transformations, leading to ambiguities in  interpreting the underlying quark configuration from the structure of the currents \cite{Zhang:2006xp}.

 QCD sum-rules have also been used to study mixed interpretations of the X(3872). If mixed molecular and charmonium currents are used, the mass prediction decreases to about $3.8\,{\rm GeV}$ \cite{Matheus:2009vq}.
From the QCD sum-rules  perspective, the four-quark state (or $\bar D D^*$ molecular state) may  couple more strongly to the charmonium hybrid than the $c\bar c$ charmonium,
because the  latter is $\alpha_s$ suppressed in perturbation theory ({\it i.e.,} the leading-order correction is chirally suppressed).  The purpose of this paper is to explore whether the X(3872)  can be described through a mixture of hybrid and molecular currents. 
  Although naive considerations seem to preclude such a mixing scenario because
 the expected masses of both the four-quark state and hybrid from
QCD sum-rules 
are widely separated
\cite{Matheus:2006xi,Harnett:2012gs,Chen:2012pe,Chen:2010ze,Nielsen:2008},
 our detailed analysis shows that a
 proper mixing current may lower the predicted mass compared to the pure charmonium hybrid. Varying the mixing parameter,
we can find a mass  near X(3872),  suggesting that the mixing between the  molecular
 and hybrid components lowers the energy
 to form a more stable state.  The resulting sum-rule is also better behaved  than the pure molecular case because we are able to find a sum-rule window of validity for a smaller continuum contribution (30\% as opposed to 50\%).

 Our paper is organized as follows:
Section~\ref{theCorrelationFunction}  contains the field-theoretical calculation of the mixed correlator of molecular and hybrid charmonium currents,  Section~\ref{theAnalysis} presents the Laplace QCD sum-rule analysis, and concluding remarks
are in Section~\ref{theConclusion}.

\section{Correlation Functions}
\label{theCorrelationFunction}

The fundamental two-point correlation functions of interest in the QCD sum-rule mixing analysis are given by
\begin{gather}
 \Pi^{\rm ij}_{\mu\nu}\left(q\right) = i \int d^4 x \, e^{i q \cdot x} \langle 0 | T\left[\right. J_\mu^{\rm i}\left(x\right) J_\nu^{\rm j}\left(0\right) \left.\right] | 0 \rangle \,,
\label{correlation_functions}
\end{gather}
where the indices $\left\{{\rm i}\,,{\rm j}\right\} \in \left\{{\rm h}\,,{\rm m}\right\}$ denote the either the hybrid or molecular currents. The hybrid diagonal correlation function $\left({\rm i}={\rm j}={\rm h}\right)$ was studied in Ref.~\cite{Harnett:2012gs} using the current
\begin{gather}
 J_\mu^h = \frac{1}{2} g \bar{c} \gamma^\nu \lambda^a \tilde{G}_{\mu\nu}^a c \,, \quad \tilde{G}_{\mu\nu}^a = \frac{1}{2} \epsilon_{\mu\nu\alpha\beta} \tilde{G}^{\alpha\beta}_a \,,
\label{hybrid_current}
\end{gather}
where $c$ denotes a charm quark field. The molecular diagonal correlation function $\left({\rm i}={\rm j}={\rm m}\right)$ was investigated in Ref.~\cite{Nielsen:2008}, using the current
\begin{gather}
 J_\nu^m = \frac{1}{\sqrt{2}}\left(\bar{q}_a\gamma_5 c_a \bar{c}_b \gamma_\nu q_b - \bar{q}_a\gamma_\nu c_a \bar{c}_b \gamma_5 q_b \right) \,,
 \label{molecular_current}
\end{gather}
where $q$ denotes a light quark field and $a\,,b$ are spinor indices.
As mentioned earlier, the mass predictions emerging from tetraquark and molecular currents are very similar
\cite{Narison:2010pd}; we have focussed on the molecular currents because the analysis can be contrasted with the scenario of mixed molecular and charmonium currents \cite{Matheus:2009vq}.
In this work we consider mixed molecular/hybrid currents
\begin{equation}
J_\nu^\xi=\sqrt{1-\xi^2}J_\nu^m+\xi \sigma J_\nu^h\,.
\label{j_xi}
\end{equation}
where $\xi$ is a dimensionless parameter that will be varied in the Section~\ref{theAnalysis} analysis and $\sigma$ is a mass scale that accounts for the different mass dimensions of the hybrid and molecular currents.
With no loss of generality, we set $\sigma=1\,{\rm GeV}$ comparable to the $\overline{\rm MS}$ charm mass; changes in $\sigma$ are compensated by a change in $\xi$.
 The correlation function
associated with this current
\begin{gather}
 \Pi^{\xi}_{\mu\nu}\left(q\right) = i \int d^4 x \, e^{i q \cdot x} \langle 0 | T\left[\right. J_\mu^{\xi}\left(x\right) J_\nu^{\xi}\left(0\right) \left.\right] | 0 \rangle
\label{mix_correlation_function}
\end{gather}
is simply a linear combination of the Eq.~\eqref{correlation_functions} correlators.  Thus the remaining quantity needed for the mixing analysis is the non-diagonal correlation function given by
\begin{gather}
 \Pi^{\rm hm}_{\mu\nu}\left(q\right) = i \int d^4 x \, e^{i q \cdot x} \langle 0 | T\left[\right. J_\mu^{\rm h}\left(x\right) J_\nu^{\rm m}\left(0\right) \left.\right] | 0 \rangle \,.
\label{non-diagonal_correlation_function}
\end{gather}
Only the transverse part of the correlation function \eqref{non-diagonal_correlation_function} couples to hadronic states with $J^{PC}=1^{++}$ that are mixtures of hybrid and molecular states, and hence we calculate
\begin{gather}
 \Pi_{\rm hm}\left(q\right) = \frac{1}{d-1} \Pi^{\mu\nu}_{\rm hm}\left(q\right)  \left(g_{\mu\nu}-\frac{q_\mu q_\nu}{q^2} \right) \,,
\end{gather}
where $d$ denotes the number of spacetime dimensions.

\begin{figure}[hbt]
\centering
\includegraphics[scale=0.35]{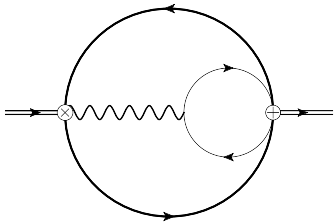}
\caption{Feynman diagram for the leading order perturbative contribution to the non-diagonal correlation function. The hybrid current is represented by the $\otimes$ symbol, the molecular current by the $\oplus$ symbol, bold quark lines represent charm quark propagators, thin quark lines represent light quark propagators, and the wavy line represents a gluon propagator. This and all subsequent diagrams were created with Jaxodraw \cite{Binosi:2003yf}.}
\label{pert_fig}
\end{figure}

We begin by calculating the leading-order perturbative contribution to the non-diagonal correlation function \eqref{non-diagonal_correlation_function}, which is depicted in Fig.~\ref{pert_fig}.\footnote{A related perturbative diagram where the gluon connects to the heavy quark line is trivially zero because of the massless (light-quark) tadpole.} As in Ref.~\cite{Harnett:2012gs} we have chosen to calculate the entire correlation function, rather than just the imaginary part. This approach provides greater clarity in the renormalization issues detailed below, and as argued in \cite{Harnett:2012gs},  justifies any limiting procedures that may be needed in forming the Laplace sum-rule.
We first note that Fig.~\ref{pert_fig} would appear to require the calculation of non-trivial massive three-loop momentum integrals. However, the light quark loop in Fig.~\ref{pert_fig} can be integrated immediately and the remaining two-loop integrals are tabulated in Ref.~\cite{Broadhurst:1993mw}, rendering the calculation tractable. Integrals with irreducible tensor structures can be calculated using the method described in Refs.~\cite{Tarasov:1996br,Tarasov:1997kx} whereby tensor integrals in $d$
dimensions are related to scalar integrals in $d+N$ dimensions. In order to use this approach, the integrals given in Ref.~\cite{Broadhurst:1993mw} must be expressed in an arbitrary number of dimensions. The relevant result is
\begin{gather}
\frac{1}{\mu^{2(d-4)}} \int \frac{d^dk_1}{(2\pi)^d} \frac{d^dk_2}{(2\pi)^d} \frac{1}{\left(k_2^2-m^2+i\,0^+\right)^\alpha\left[\right.\left(k_2-q\right)^2-m^2+i\,0^+\left.\right]^\beta \left[\left(k_1-k_2\right)^2+i\,0^+\right]^\gamma}
\\
\begin{split}
=-\frac{1}{(4\pi)^4}  \left(-m^2\right)^{(4-\alpha-\beta-\gamma)}  \left[\frac{m^2}{4\pi\mu^2}\right]^{(d-4)} &
\frac{\Gamma\left(\alpha+\beta+\gamma-d\right)\Gamma\left(\frac{d}{2}-\gamma\right)\Gamma\left(\beta+\gamma-\frac{d}{2}\right)\Gamma\left(\alpha+\gamma-\frac{d}{2}\right)}{\Gamma\left(\alpha\right)\Gamma\left(\beta\right)\Gamma\left(\frac{d}{2}\right)\Gamma\left(\alpha+\beta+2\gamma-d\right)}
\\
&\phantom{}_4 F_{3}
\left[
\begin{array}{c|}
\gamma \,,  \alpha+\beta+\gamma-d \,,  \beta+\gamma-\frac{d}{2} \,,   \alpha+\gamma-\frac{d}{2} \\
 \frac{d}{2} \,,  \gamma+\frac{1}{2}\left(\alpha+\beta-d\right) \,,  \gamma+\frac{1}{2}\left(\alpha+\beta-d+1\right)
\end{array} \, \frac{q^2}{4m^2}  \right] \,,
\end{split}
\label{TJI_integral_result}
\end{gather}
where $\mu$ is the renormalization scale, $d$ is the number of spacetime dimensions, $m$ is the charm quark mass, $q^2$ is the external momentum and $\phantom{}_4 F_{3}$ denotes the generalized hypergeometric function (we use the hypergeometric function conventions of Ref.~\cite{Bateman:1953}). Note that the we have modified the $d=4-2\epsilon$ normalization of the loop integrals that was used in Ref.~\cite{Broadhurst:1993mw} to facilitate working in higher dimensions.\footnote{Ref.~\cite{DavydychevTalk} also gives a result for the integral \eqref{TJI_integral_result} which is consistent with our conventions.} The perturbative contribution to the non-diagonal correlator can now be calculated using \eqref{TJI_integral_result} combined with the methods of \cite{Tarasov:1996br,Tarasov:1997kx}. Using the $\overline{\rm MS}$ scheme and working in $d=4+2\epsilon$ dimensions, we find
\begin{gather}
\begin{split}
\Pi_{\rm hm}^{\rm pert}\left(z\right) = -\frac{m^7\,\alpha}{24\sqrt{2}\pi^5}&\Biggl[\Biggr. 3 {\rm Li}_3\left[1-2z-2i\sqrt{z\left(1-z\right)} \right] + 6i {\rm Li}_2\left[1-2z-2i\sqrt{z\left(1-z\right)} \right]\sin^{-1}{\left(\sqrt{z}\right)}
\\
&+\frac{1}{192z^2} \left(896z^4-1280z^3+96z^2\left(12\log{\left[2\right]}-1\right)+80z+7\right)\left[\sin^{-1}{\left(\sqrt{z}\right)}\right]^2
\\
&-\frac{1}{1440z} \left(2340z^5+7568z^4-26568z^3+15426z^2+1165z+105\right) \frac{ \sin^{-1}{\left(\sqrt{z}\right)} }{\sqrt{z\left(1-z\right)} }
\\
&+2i\left[\sin^{-1}{\left(\sqrt{z}\right)}\right]^3+6\log\left[z+i\sqrt{z\left(1-z\right)}\right]\left[\sin^{-1}{\left(\sqrt{z}\right)}\right]^2 +\frac{7}{192z} \Biggl.\Biggr]
\,, \quad z = \frac{q^2}{4m^2} \,,
\end{split}
\label{perturbative_correlator_result}
\end{gather}
where ${\rm Li}_3$ and ${\rm Li}_2$ denote the trilogarithm and dilogarithm functions, respectively~\cite{LewinPolylogarithms}. Terms corresponding to dispersion relation subtraction constants have been omitted, and the coupling $\alpha$ and charm quark mass $m$ are implicitly functions of the renormalization scale $\mu$. Although \eqref{perturbative_correlator_result} superficially appears to be singular at $z=0$, it is in fact well-defined at this point. The imaginary part can be determined through analytic continuation of the functions in \eqref{perturbative_correlator_result}, yielding
\begin{gather}
\begin{split}
{\rm Im}\Pi_{\rm hm}^{\rm pert}\left(z\right) = -\frac{m^7\,\alpha}{48\sqrt{2}\pi^4}  \Biggl[ & 6\left({\rm Li_2}\left[-\left(\sqrt{z}+\sqrt{z-1}\right)^2\right]+\log^2{\left[\sqrt{z}+\sqrt{z-1}\right]}+\log{\left[\sqrt{z}+\sqrt{z-1}\right]}\log{\left[z\right]}\right) \Biggr.
\\
&+ \frac{1}{96z^2}\left(896z^4-1280z^3+96z^2\left(12\log{\left[2\right]}-1\right)+80z+7\right)\log{\left[\sqrt{z}+\sqrt{z-1}\right]}
\\
&-\frac{1}{1440z}\left(2340z^5+7568z^4-26568z^3+15426z^2+1165z+105\right)\frac{1}{\sqrt{z\left(z-1\right)}}
\\
&+\frac{\pi^2}{2} \Biggl. \Biggr] \,, \quad z > 1 \,.
\end{split}
\label{perturbative_imaginary_part}
\end{gather}

Now we consider contributions from the QCD condensates. These involve momentum integrals that are at most two-loop which can be dealt with using the Mathematica package Tarcer~\cite{Mertig:1998vk}. This package implements the recurrence relations developed in Refs.~\cite{Tarasov:1996br,Tarasov:1997kx} to reduce a large set of two-loop momentum integrals to a much smaller set. This small set includes  integrals that are given by \eqref{TJI_integral_result} above, as well as two additional integrals that are tabulated in Refs.~\cite{Boos:1990rg,Davydychev:1990cq}. The epsilon expansions of all of these integrals can be easily performed using the Mathematica package HypExp~\cite{Huber:2005yg,Huber:2007dx}.

\begin{figure}[hbt]
\centering
\includegraphics[scale=0.35]{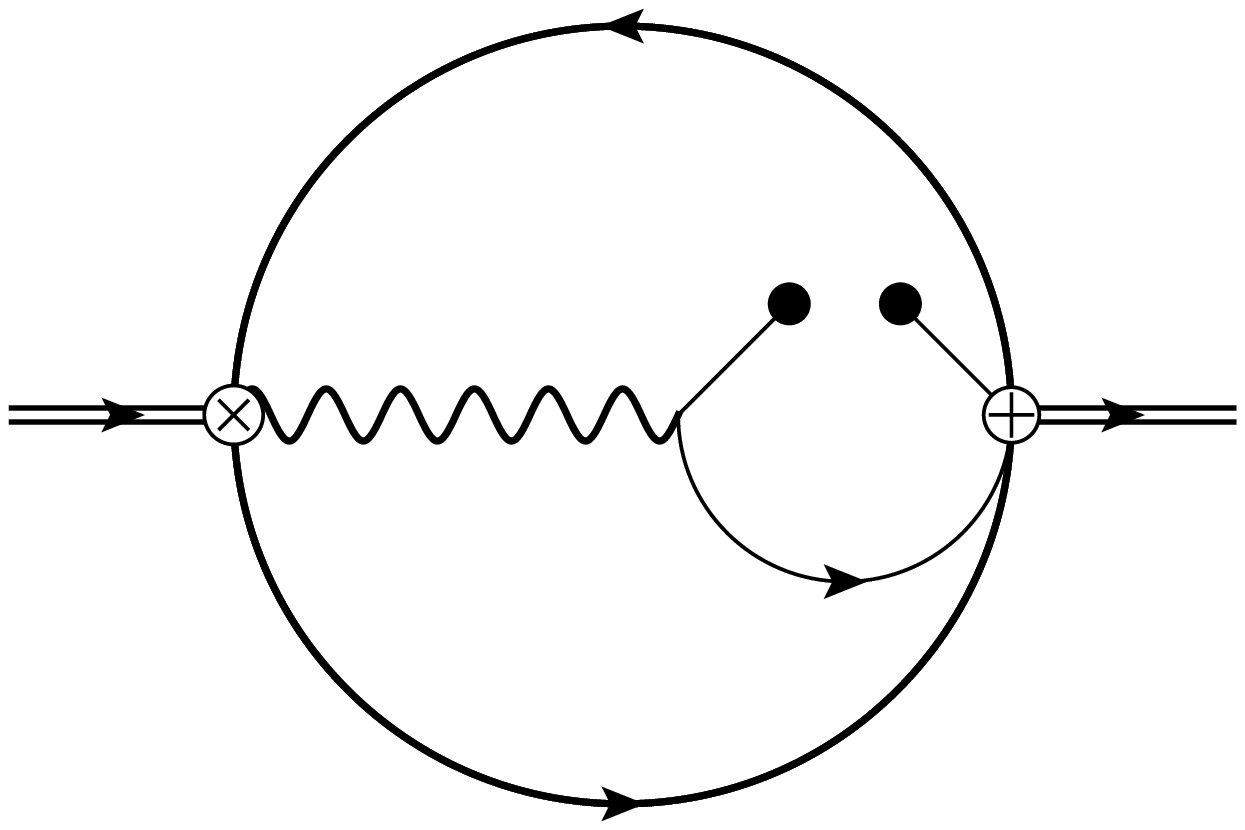}
\includegraphics[scale=0.35]{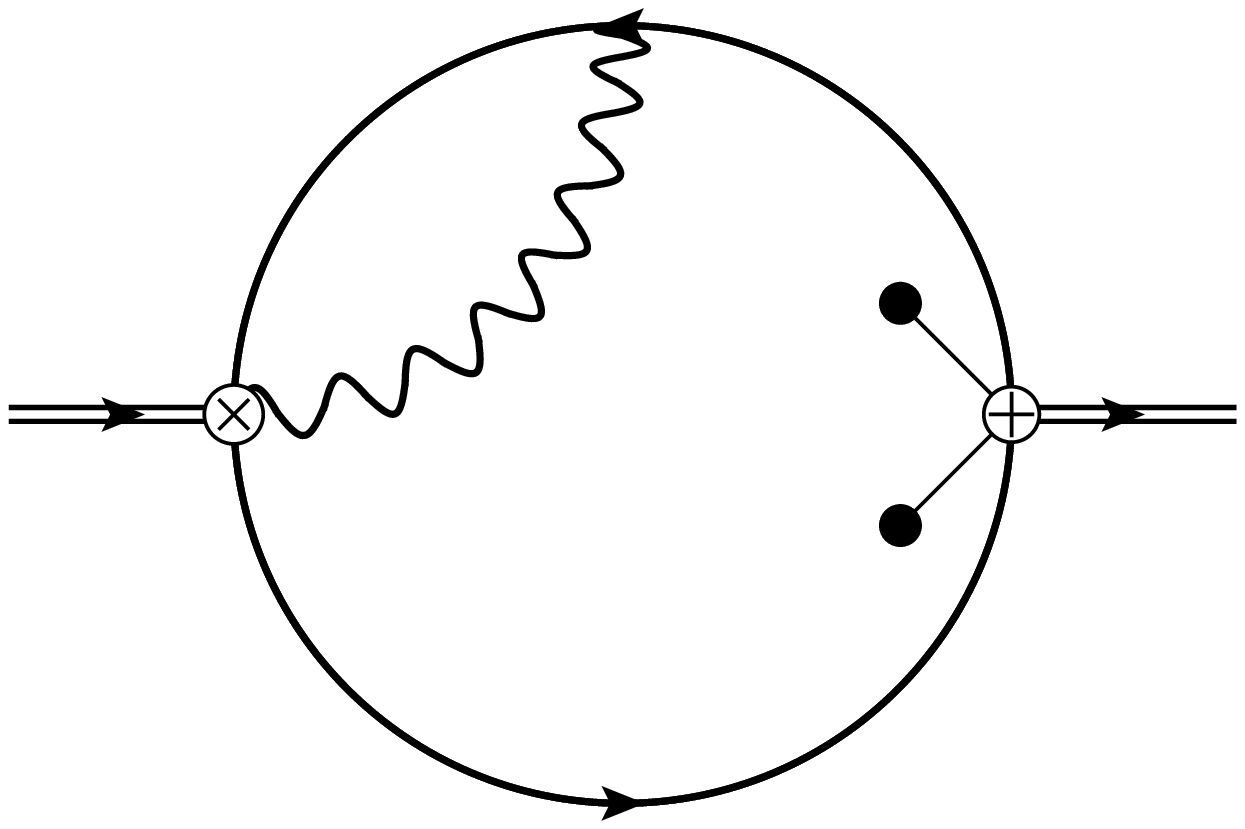}
\caption{Feynman diagrams for the leading order contributions of the (light) quark condensate to the non-diagonal correlation function. The diagram on the left corresponds to $\Pi^{\rm qq1}_{\rm hm}$ and the diagram on the right corresponds to $\Pi^{\rm qq2}_{\rm hm}$. Diagrams related by symmetry are not shown. All notations are identical to those in Fig.~\ref{pert_fig}.}
\label{qq_fig}
\end{figure}

\begin{figure}[hbt]
\centering
\includegraphics[scale=0.35]{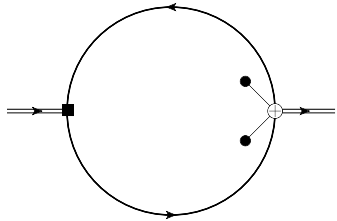}
\caption{Feynman diagram representing the contribution to the non-diagonal correlation function due to the renormalization of the hybrid current. The black box represents an insertion of the lower dimensional operator $\mathcal{O}_\mu$ that mixes with the hybrid current. All other notations are identical to those in Fig.~\ref{pert_fig}.}
\label{renorm_fig}
\end{figure}

We first consider the quark condensate, for which there are two distinct contributions as depicted in Fig.~\ref{qq_fig}. For convenience in what follows we define
\begin{gather}
w=\sqrt{\frac{z}{1-z}} \,,
\label{w_of_z}
\end{gather}
where $z$ is as defined in \eqref{perturbative_correlator_result}.  For the left diagram in Fig.~\ref{qq_fig}, we find
 \begin{gather}
\begin{split}
 \Pi^{\rm qq1}_{\rm hm}\left(z\right) = -\frac{m^4 \alpha \langle \bar{q}q \rangle}{1152\sqrt{2}\pi^3z^2} \Biggl[  4i & \left(  16z^4-104z^3+  46z^2+51z-9\right) w\log{\left[\frac{i-w}{i+w}\right]} \Biggr.
 \\
 &\Biggl. +3\left(3-16z+48z^2\right)\log^2{\left[\frac{i-w}{i+w}\right]} \Biggr] \,.
 \end{split}
\label{quark_condensate_1_correlator_result}
\end{gather}
The imaginary part of \eqref{quark_condensate_1_correlator_result} is
\begin{gather}
\begin{split}
 {\rm Im}\Pi^{\rm qq1}_{\rm hm}\left(z\right) = \frac{m^4 \alpha \langle \bar{q}q \rangle}{96\sqrt{2}\pi^2} \Biggl[
 \frac{1}{z^2}&\left(48z^2-16z+3\right) \log{\left[\sqrt{z}+\sqrt{z-1}\right]} \Biggr.
 \\
 &\Biggl.+\frac{1}{3z}\left(16z^4-104z^3+46z^2+51z-9\right)\frac{1}{\sqrt{z\left(z-1\right)}} \Biggr] \,, \quad z > 1 \,.
\end{split}
\label{quark_condensate_1_imaginary_part}
\end{gather}
For the right diagram in Fig.~\ref{qq_fig}, we find
\begin{gather}
\begin{split}
 \Pi^{\rm qq2}_{\rm hm, bare}\left(z\right) = \frac{m^4\alpha \langle \bar{q}q \rangle}{162\sqrt{2}\pi^3} \Biggl[ &
 \frac{30i(z-1)}{w}\log{\left[\frac{i-w}{i+w}\right]} \frac{1}{\epsilon}
 + \frac{30i(z-1)}{w}\left(  {\rm Li}_2\left[\frac{1-iw}{2}\right] - {\rm Li}_2\left[\frac{1+iw}{2}\right] \right.  \Biggr.
 \\
 &\left. +\frac{1}{2}\log^2{\left[1-iw\right]}-\frac{1}{2}\log^2{\left[1+iw\right]} - \log{\left[2\right]}\log{\left[\frac{1-iw}{1+iw}\right]} \right)
 \\
 &-\frac{i}{4zw} \left( 96z^3 +140z^2 +16z -15 -240z(z-1)\log{\left[\frac{m^2}{\mu^2}\right]} \right)\log{\left[\frac{i-w}{i+w}\right]}
 \\
 &\Biggl.-\frac{3}{16z^2}\left(80z^3-24z^2-6z+5\right)\log^2{\left[\frac{i-w}{i+w}\right]} \Biggr]\,.
\end{split}
\label{quark_condensate_2_bare_correlator_result}
\end{gather}
Notice that the first term in \eqref{quark_condensate_2_bare_correlator_result} contains a non-local divergence. This term is problematic since it cannot be removed through dispersion relation subtraction constants or application of the Borel transform when the sum rules are formulated, nor can it be removed through a multiplicative renormalization. Similar to the mixed scalar gluonic and quark currents \cite{Harnett:2008cw}, the origin of this divergence is the renormalization-induced mixing  of the composite operator \eqref{hybrid_current}
 with operators of lower mass dimension. This is merely a field theoretical counterpart to the idea that since the $J^{PC}=1^{++}$ hybrid is non-exotic, it can and will mix with conventional charmonium states with the same quantum numbers. This is in contrast to hybrids with exotic $J^{PC}$, which do not mix with conventional charmonium states (see \textit{e.g.}, Ref.~\cite{Jin:2002rw}). The renormalized $1^{++}$ hybrid current can be expressed as
\begin{gather}
\left[J^h_\mu\right]_R = Z_1 \left[J^h_\mu\right]_B + Z_2 m^2 \left[\mathcal{O}_\mu\right]_B + \ldots \,, \quad
Z_1 = 1+\frac{\alpha}{\pi}\frac{Z_{h1}}{\epsilon} \,, \quad Z_2 = \frac{\alpha}{\pi}\frac{Z_{h2}}{\epsilon} \,,
\label{renormalized_hybrid_current}
\end{gather}
where $m$ denotes the charm quark mass, the subscripts $R$ and $B$ represent renormalized and bare quantities, respectively, and the ellipses in \eqref{renormalized_hybrid_current} are to indicate that additional lower dimensional operators may be present that would only contribute the the mixed-correlator \eqref{non-diagonal_correlation_function} at higher-loop level. The first term in \eqref{renormalized_hybrid_current} corresponds to the multiplicative renormalization of the hybrid current, which could be relevant for higher-order studies of hybrid-molecular state mixing, but is irrelevant to us at present because it represents a higher-loop effect. The composite operator in \eqref{renormalized_hybrid_current} is given by
\begin{gather}
\mathcal{O}_\mu = \bar{c} \Gamma_\mu c \,, \quad \Gamma_\mu = \epsilon_{\mu\nu\alpha\beta} \left( \gamma^\nu \sigma^{\alpha\beta} + \gamma^\alpha \sigma^{\beta\nu}-\gamma^\beta \sigma^{\alpha\nu} \right) \,,
\label{operator_O}
\end{gather}
where $\sigma^{\alpha\beta} = \frac{i}{2}\left[\gamma^\alpha\,,\gamma^\beta\right]$ and $c$ denotes a charm quark field. Therefore, there is a renormalization induced contribution to the quark condensate, arising from the composite operator $\mathcal{O}_\mu$ in \eqref{renormalized_hybrid_current}. This contribution is denoted as $\Pi^{\rm qq2}_{\rm hm, opmix}$ and is represented in Fig.~\ref{renorm_fig}.\footnote{The renormalization-induced perturbative diagram is trivially zero because of the massless (light-quark)  tadpole. }

To our knowledge, the renormalization properties of the $1^{++}$ hybrid current have never been studied. As such a full study is beyond the scope of the present work, we simply tune the renormalization factor $Z_{h2}$ such that the non-local divergence in \eqref{quark_condensate_2_bare_correlator_result} is cancelled. That is, we require that the sum
\begin{gather}
\Pi^{\rm qq2}_{\rm hm, renorm} = \Pi^{\rm qq2}_{\rm hm, bare} + \Pi^{\rm qq2}_{\rm hm, opmix}
\label{renormalization_condition}
\end{gather}
has no non-local divergences. Doing so, we find
\begin{gather}
Z_{h2} = -\frac{10}{243}\frac{\alpha}{\pi} \frac{1}{\epsilon} \,,
\label{renormalization_factor}
\end{gather}
and the renormalized induced contribution to the quark condensate is
\begin{gather}
\begin{split}
 \Pi^{\rm qq2}_{\rm hm, opmix}\left(z\right) = -\frac{m^4\alpha \langle \bar{q}q \rangle}{162\sqrt{2}\pi^3} \Biggl[ &
 \frac{30i(z-1)}{w}\log{\left[\frac{i-w}{i+w}\right]} \frac{1}{\epsilon}
 + \frac{30i(z-1)}{w}\left\{  {\rm Li}_2\left[\frac{1-iw}{2}\right] - {\rm Li}_2\left[\frac{1+iw}{2}\right] \right.  \Biggr.
 \\
 & -\frac{1}{2}\log^2{\left[1+iw\right]} + \frac{1}{2}\log{\left[\frac{1-iw}{4}\right]}\log{\left[\frac{1-iw}{1+iw}\right]}
 \\
 &\Biggl.\left.+ \frac{1}{2}\log{\left[1-iw\right]}\log{\left[1+iw\right]} +\frac{i(z-1)}{3w} \left( 5+3\log{\left[\frac{m^2}{\mu^2}\right]} \right)\log{\left[\frac{i-w}{i+w}\right]}\right\} \Biggr] \,.
\end{split}
\label{operator_mixing_result}
\end{gather}
Comparing \eqref{quark_condensate_2_bare_correlator_result} and \eqref{operator_mixing_result}, it is clear that the non-local divergence will be eliminated. The renormalized quark condensate contribution is then
\begin{gather}
\begin{split}
 \Pi^{\rm qq2}_{\rm hm, renorm} \left(z\right) = -\frac{m^4 \alpha \langle \bar{q}q \rangle}{162\sqrt{2}\pi^3} \Biggl[ &  \frac{i}{4zw}  \left(  96z^3 +340z^2 -184z -15 -120z(z-1)\log{\left[\frac{m^2}{\mu^2}\right]} \right) \log{\left[\frac{i-w}{i+w}\right]}  \Biggr.
 \\
 & \Biggl. +\frac{3}{16z^2}\left(80z^3-24z^2-6z+5\right)\log^2{\left[\frac{i-w}{i+w}\right]} \Biggr] \,.
\end{split}
\label{quark_condensate_2_renormalized_correlator_result}
\end{gather}
The imaginary part of \eqref{quark_condensate_2_renormalized_correlator_result} can now be easily extracted, yielding
\begin{gather}
\begin{split}
 {\rm Im}\Pi^{\rm qq2}_{\rm hm, renorm}\left(z\right) = \frac{m^4 \alpha \langle \bar{q}q \rangle}{216\sqrt{2}\pi^2} \Biggl[ &
 \frac{1}{z^2}\left(80z^3-24z^2-6z+5\right)  \log{\left[\sqrt{z}+\sqrt{z-1}\right]} \Biggr.
 \\
 &\Biggl. -\frac{1}{3z}\left(  96z^3 +340z^2 -184z -15 -120z(z-1)\log{\left[\frac{m^2}{\mu^2}\right]} \right) \sqrt{\frac{z-1}{z}} \Biggr] \,, \quad z>1 \,.
\end{split}
\label{quark_condensate_2_imaginary_part}
\end{gather}

\begin{figure}[hbt]
\centering
\includegraphics[scale=0.35]{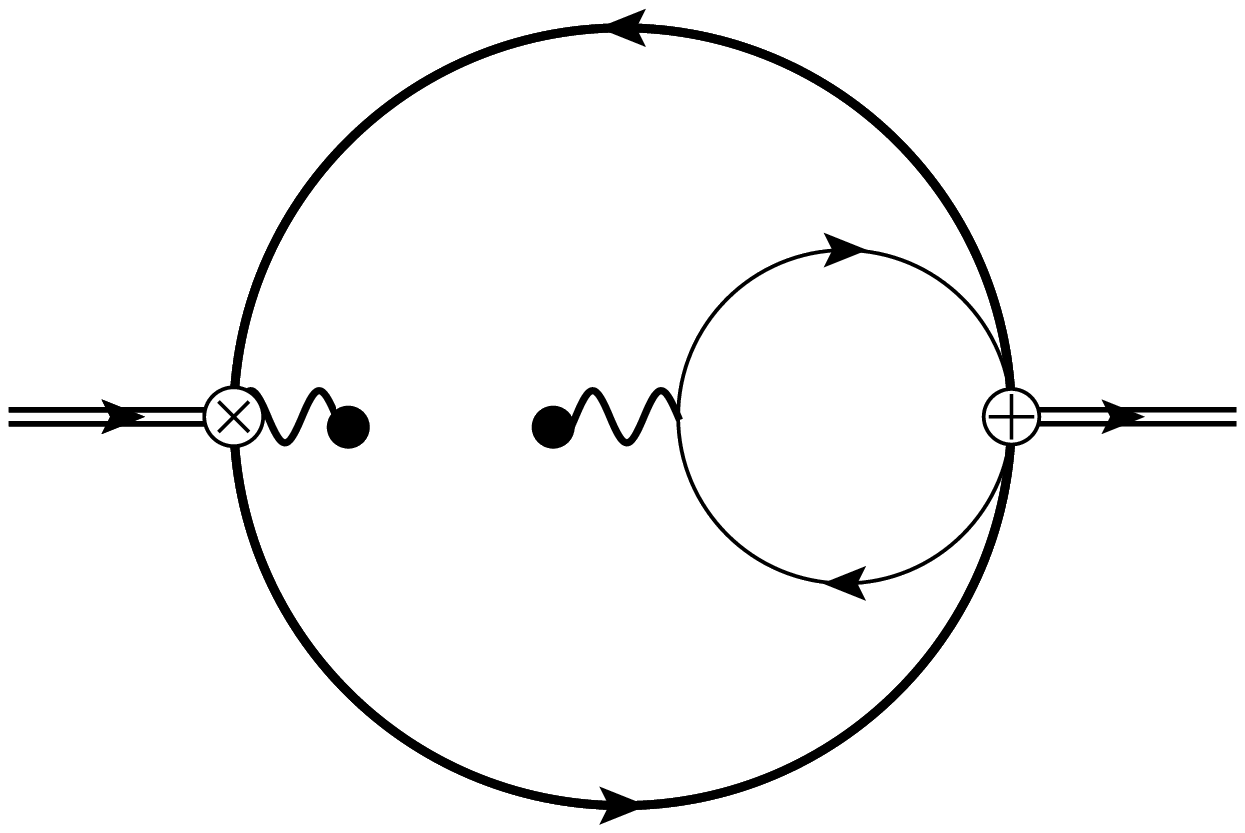}
\caption{Feynman diagram for the leading order contribution of the gluon condensate to the non-diagonal correlation function. All notations are identical to those in Fig.~\ref{pert_fig}.}
\label{g2_fig}
\end{figure}

Note that the gluon condensate, as would be obtained from Fig.~\ref{g2_fig} is chirally suppressed by the light quark loop and hence has a negligible effect on the analysis.\footnote{A related gluon condensate diagram where the gluon connects to the heavy quark line is trivially zero because of the massless (light-quark) tadpole.} This same chiral suppression of the gluon condensate occurs in the mixed correlator of scalar (light) quark and gluonic currents~\cite{Harnett:2008cw}.

\begin{figure}[hbt]
\centering
\includegraphics[scale=0.35]{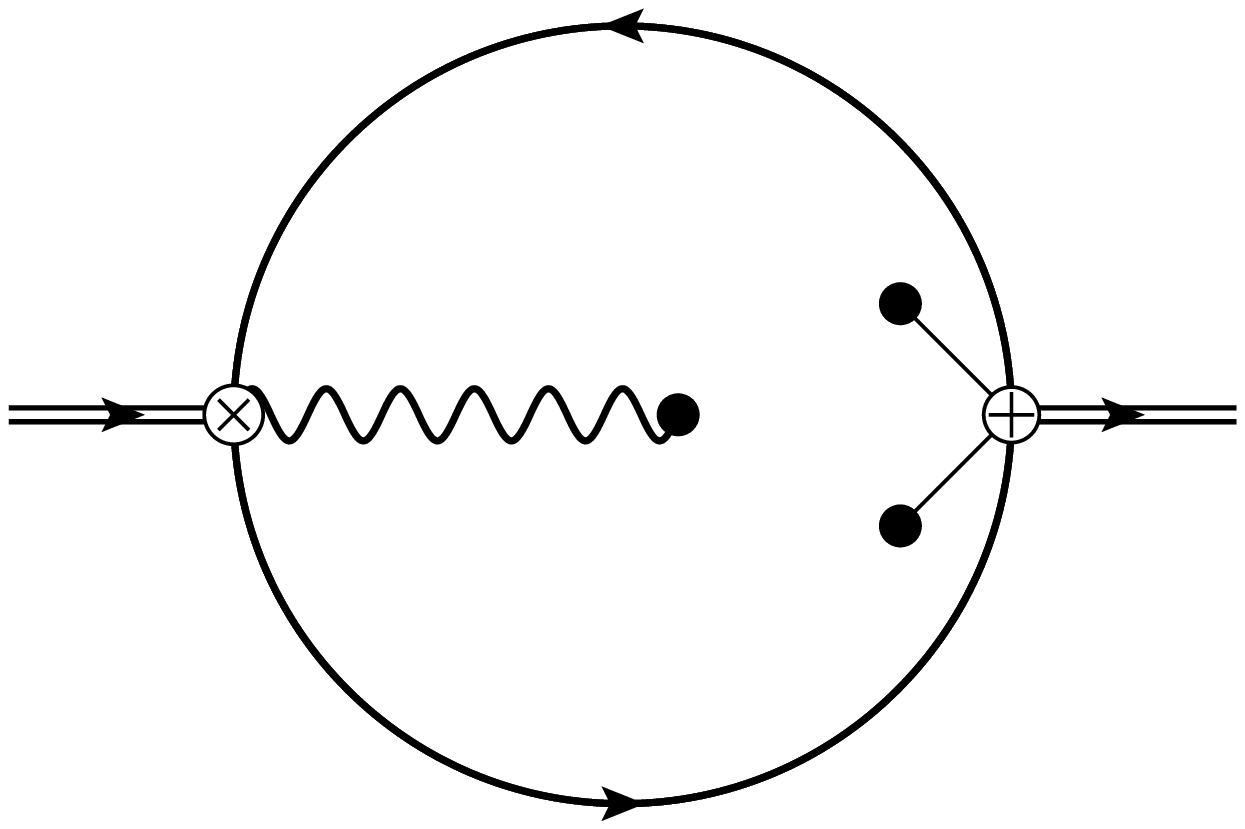}
\caption{Feynman diagram for the leading order contribution of the mixed condensate to the non-diagonal correlation function. All notations are identical to those in Fig.~\ref{pert_fig}.}
\label{mixed_fig}
\end{figure}

Finally we consider contributions from the mixed condensate, which is depicted in Fig.~\ref{mixed_fig}. For this contribution we find
\begin{gather}
\begin{split}
 \Pi^{\rm qGq}_{\rm hm}\left(z\right) = \frac{im^2 \langle \bar{q}\sigma G q \rangle}{72\sqrt{2}\pi^2} \frac{w}{z} \left(2z^2-z-1\right)\log{\left[\frac{i-w}{i+w}\right]} \,,
\end{split}
\label{mixed_condensate_result}
\end{gather}
where we define $\langle \bar{q}\sigma G q \rangle = \langle g \bar{q} \frac{\lambda^a}{2} \sigma_{\mu\nu} G^a_{\mu\nu} q \rangle$. The corresponding imaginary part is
\begin{gather}
\begin{split}
 {\rm Im}\Pi^{\rm qGq}_{\rm hm}\left(z\right) = -\frac{m^2 \langle \bar{q}\sigma G q \rangle }{72\sqrt{2}\pi} (1+2z) \frac{\sqrt{z-1}}{\sqrt{z}} \,, \quad z>1 \,.
\end{split}
\label{mixed_condensate_imaginary_part}
\end{gather}

\section{QCD Laplace Sum-Rule Analysis}
\label{theAnalysis}

Utilizing the results given above for the non-diagonal hybrid-molecular correlation function $\Pi_{\rm hm}$, along with the results from Refs.~\cite{Harnett:2012gs,Nielsen:2008} for the diagonal hybrid correlation function  $\Pi_{\rm hh}$ and molecular correlation function $\Pi_{\rm mm}$, we now perform the QCD Laplace sum-rules analysis of mixing between $J^{PC}=1^{++}$ molecular and hybrid charmonium. We do not review the QCD Laplace sum-rules methodology here, but the reader is directed to the original papers~\cite{Shifman:1978bx,Shifman:1978by} and, for example, reviews given in Refs.~\cite{Reinders:1984sr,Narison:2002pw}. Invoking the standard resonance plus continuum model for the hadronic spectral function, the Laplace sum-rules take the form
\begin{equation}
 {\cal R}_{k}\left(\tau,s_0\right)  = \frac{1}{\pi}\int_{t_0}^{\infty} t^k
   \exp\left[ -t\tau\right] \rho^{\rm had}\left(t\right)\; dt \,,
\label{final_laplace}
\end{equation}
where $t_0$ is the hadronic threshold. The left hand side of \eqref{final_laplace} is given by
\begin{equation}
{\cal R}_k\left(\tau,s_0\right)\equiv\frac{1}{\tau}\hat B\left[\left(-1\right)^k Q^{2k}\Pi\left(Q^2\right)\right] -  \frac{1}{\pi}  \int_{s_0}^{\infty} t^k
   \exp \left[-t\tau  \right]  {\rm Im} \Pi\left(t\right)\; dt
\label{laplace}
\end{equation}
where $s_0$ is the continuum threshold for the hadronic spectral function $\rho^{\rm had}\left(t\right)$, $Q^2=-q^2$ is the Euclidean momentum, and $\hat B$ is the Borel transform operator. For our purposes $\Pi\left(Q^2\right)$ in \eqref{laplace} corresponds to either the non-diagonal correlator $\Pi_{\rm hm}$, or the diagonal correlators $\Pi_{\rm hh}$ and $\Pi_{\rm mm}$.

We now construct the non-diagonal hybrid-molecular sum-rules. Using the results obtained above for the perturbative \eqref{perturbative_imaginary_part}, quark condensate \eqref{quark_condensate_1_imaginary_part},\eqref{quark_condensate_2_imaginary_part}, and mixed condensate \eqref{mixed_condensate_imaginary_part} contributions to the non-diagonal correlation function, the QCD Laplace sum-rules take the form
\begin{gather}
\begin{split}
{\cal R}_0^{\rm hm}\left(\tau,s_0\right)=\frac{4m^2}{\pi}\int_1^{s_0/4m^2} \Biggl[ {\rm Im}\Pi_{hm}^{\rm pert}\left(4m^2 x\right)&+{\rm Im}\Pi_{\rm hm}^{\rm qq1}\left(4m^2 x\right) +{\rm Im}\Pi_{\rm hm, renorm}^{\rm qq2}\left(4m^2 x\right) \Biggr.
\\
&+\Biggl.{\rm Im}\Pi^{\rm qGq}_{\rm hm}\left(4m^2 x\right)\Biggr]\exp{\left(-4m^2\tau x\right)}\,dx \,,
\label{hybrid_molecule_non_diagonal_sum_rule_0}
\end{split}
\\
{\cal R}_{1}^{\rm hm}\left(\tau,s_0\right)=-\frac{\partial}{\partial\tau}{\cal R}_{0}^{\rm hm}\left(\tau,s_0\right)\,.
\label{hybrid_molecule_non_diagonal_sum_rule_1}
\end{gather}
The mass and coupling in \eqref{hybrid_molecule_non_diagonal_sum_rule_0},\eqref{hybrid_molecule_non_diagonal_sum_rule_1} are implicitly functions of the renormalization scale $\mu$ in the $\overline{\rm MS}$-scheme, and renormalization group improvement may be implemented by setting $\mu=1/\sqrt{\tau}$ after evaluating the derivative with respect to $\tau$~\cite{Narison:1981ts} or by setting $\mu$ to the charm quark mass scale. The sum-rules for the diagonal hybrid and molecular correlation functions are given in Refs.~\cite{Harnett:2012gs,Nielsen:2008}, respectively. As these are both somewhat lengthy expressions, we do not repeat them here.
In terms of the correlation function \eqref{mix_correlation_function}  of the mixed current \eqref{j_xi}, the Laplace sum-rule is then a linear combination of the diagonal and non-diagonal expressions
\begin{equation}
{\cal R}_k^{\xi}\left(\tau,s_0\right)=\xi^2 \sigma^2{\cal R}_k^{\rm hh}\left(\tau,s_0\right)
+\left(1-\xi^2\right){\cal R}_k^{\rm mm}\left(\tau,s_0\right)+2\xi\sqrt{1-\xi^2} \sigma{\cal R}_k^{\rm hm}\left(\tau,s_0\right)\,.
\label{R_xi}
\end{equation}

If the parameter $\xi$ is chosen appropriately, then a single narrow resonance model can be used:
\begin{equation}
 \frac{1}{\pi}\rho^{\rm had}(t)=f^2\delta\left(t-M_X^2\right)\,.
 \label{narrow_res}
\end{equation}
Eq.~\eqref{final_laplace} then yields
\begin{equation}
{\cal R}_k^{\xi}\left(\tau,s_0\right)=f^2 M_X^{2k}\exp{\left(-M_X^2\tau\right)}\,,
\label{narrow_sr}
\end{equation}
from which the ground state mass $M_X$ can be determined via the ratio
\begin{equation}
M_X^2=\frac{{\cal R}_1^{\xi}\left(\tau,s_0\right)}{{\cal R}_0^{\xi}\left(\tau,s_0\right)}\,.
\label{ratio}
\end{equation}

The QCD input parameters within the sum-rules are chosen to maintain consistency with the diagonal charmonium  hybrid analysis ~\cite{Harnett:2012gs}
and the molecular analysis \cite{Nielsen:2008} (see \cite{Beringer:1900zz,Bethke:2009jm,Dosch:1988vv,Narison:2010cg} for the original sources of the parameter values):
\begin{gather}
m\left(\mu=m_c\right)=\overline m_c=\left(1.28\pm 0.02\right)\,{\rm GeV}\,,
\label{mc_mass}\\
\alpha\left(M_\tau\right)=0.33\,;
\label{running_coupling}\\
 \langle \bar{q}\sigma G q \rangle = M_0^2 \,  \langle \bar{q}q \rangle \,,~M_0^2=\left(0.8\pm0.1\right)\,{\rm GeV}^2\,,
 \label{mixed_condensate}
 \\
 \langle \bar q q\rangle=-(0.23\pm 0.03 \,{\rm GeV})^3\,,
 \label{quark_condensate}
 \\
 \langle \alpha G^2\rangle=\left(7.5\pm 2.0\right)\times 10^{-2}\,{\rm GeV^4}\,,
\label{GG_value}
\\
\langle g^3G^3\rangle=\left(8.2\pm 1.0\right){\rm GeV^2}\langle \alpha G^2\rangle\,.
\label{GGG_value}
\end{gather}

Our analysis methodology is to optimize the parameter $\xi$ to find the best agreement between $M_X$ and the  $X(3872)$.  For a given value of $\xi$ we find a sum-rule window of validity where the continuum contribution is less than 30\% (providing an upper bound on the Borel scale $M^2=1/\tau$) and where the condensate contributions are less than 10\% (providing a lower bound on the Borel scale $M$).
In this sum-rule window of validity, we require that $M_X$ has a critical point (typically a minimum) as a function of the Borel scale.
Within the range of continuum $4.1\,{\rm GeV} \lesssim \sqrt{s_0} \lesssim 5.5\,{\rm GeV}$ previously found for the pure charmonium hybrid \cite{Harnett:2012gs} and molecular \cite{Nielsen:2008} sum-rules,  $s_0$ is optimized to minimize the dependence
of $M_X$ on the Borel scale.

\begin{figure}[hbt]
\centering
\includegraphics[scale=0.6]{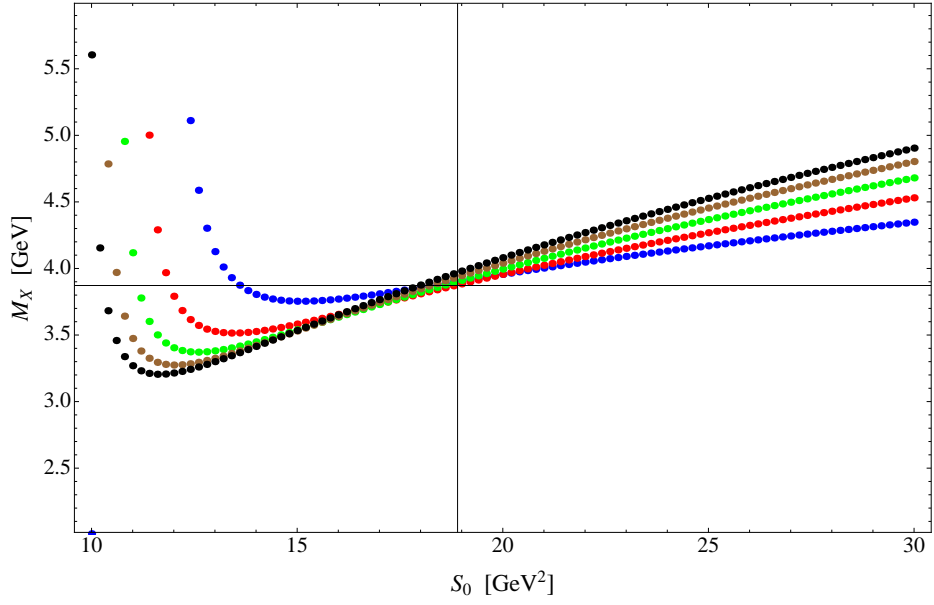}
\caption{The sum-rule prediction of $M_X$ as a function of the continuum $s_0$ for $\xi=0.002$  for Borel scales  $M$ within the  sum-rule region of validity: $M^2=\{1.6, ~1.75,~1.9,~2.05,~2.2\}\,{\rm GeV^2}$. The minimum value of $s_0$ corresponds to the pure molecular analysis \cite{Nielsen:2008} and the maximum value corresponds to the pure hybrid case
\cite{Harnett:2012gs}. The  ${\overline{\rm MS}}$ scheme has been used for the charm quark mass.
 }
\label{continuum_fig}
\end{figure}

\begin{figure}[hbt]
\centering
\includegraphics[scale=0.6]{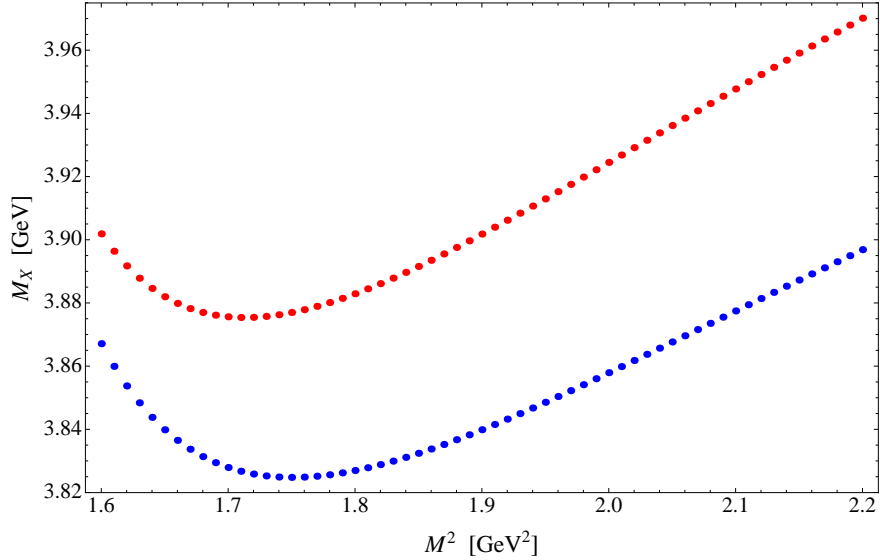}
\caption{The sum-rule prediction of $M_X$ as a function of the Borel scale $M$ for $\xi=0.002$ and the optimized continuum $s_0=19\,{\rm GeV^2}$ (red curve) and a slightly smaller value $s_0=18\,{\rm GeV^2}$ (blue curve).
The range of $M$ corresponds to the sum-rule region of validity $1.6\,{\rm GeV}^2<M^2<2.2\,{\rm GeV}^2$ for the central $s_0$ values.
The  ${\overline{\rm MS}}$ scheme has been used for the charm quark mass.} 
\label{mass_fig}
\end{figure}

As $\xi$ is increased from zero  (\textit{i.e.} the pure molecular case), $M_X$  increases until it reaches a maximum value near $\xi=0.002$, and then decreases until the sum-rule becomes unstable.
For this value $\xi=0.002$, the optimization of $s_0$  shown in Fig.~\ref{continuum_fig}  results in $\sqrt{s_0}\approx 4.3\,{\rm GeV}$, and
 Fig.~\ref{mass_fig} shows the prediction of $M_X$ corresponding to the largest mass prediction.   The critical point in  Fig.~\ref{mass_fig} provides the mass prediction $M_X=3.88\,{\rm GeV}$ in excellent correspondence with the $X(3872)$.
Although the value of $\xi$ appears to represent a small mixing between the hybrid and molecular currents, it in fact represents a significant mixing effect because the ratio of the perturbative contributions to the sum-rules is quite large: $10^3\lesssim\sigma^2{\cal R}_0^{\rm hh}/{\cal R}_0^{\rm mm}\lesssim 10^5$.   As mentioned earlier, the choice of $\sigma$ affects $\xi$;  if we rescale  the hybrid current by setting $1/\sigma \sim \sqrt{ {\cal R}_0^{\rm hh}/{\cal R}_0^{\rm mm}}$, then  $0.06 \lesssim\xi \lesssim 0.6$.
The mixing of hybrid and molecular currents seems to have an important stabilizing effect on the sum-rule analysis.  For the pure molecular case,  it is difficult to control the continuum contributions and it is necessary to accept a sum-rule analysis where the continuum reaches 50\% \cite{Matheus:2006xi,Chen:2012pe,Chen:2010ze,Nielsen:2008}.  However, in the mixed case  we are able to find a working region where the continuum contribution is less than 30\%.

\begin{figure}[hbt]
\centering
\includegraphics[scale=0.6]{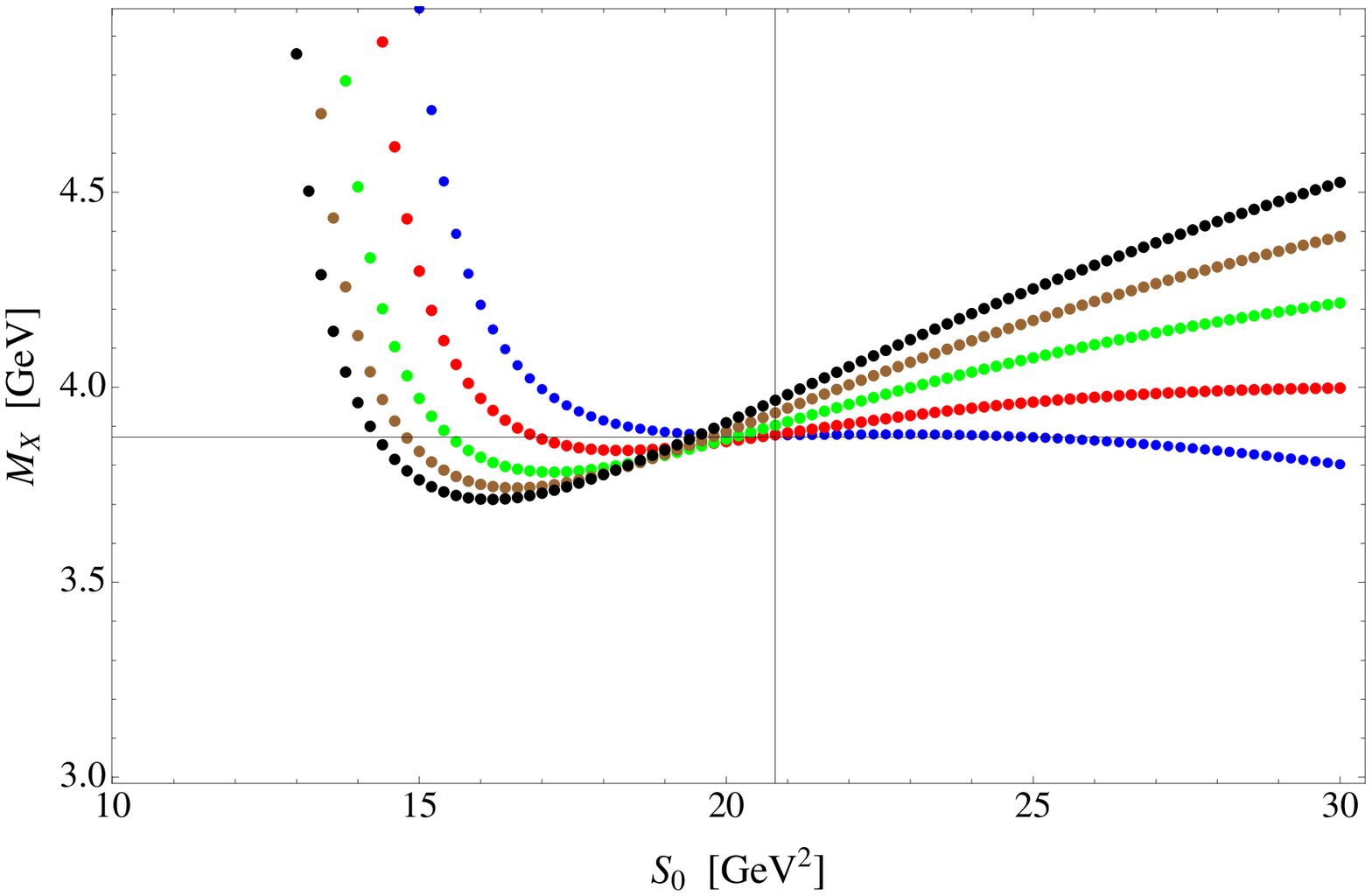}
\caption{
The sum-rule prediction of $M_X$ as a function of the continuum $s_0$ for $\xi=0.007$  for Borel scales  $M$ within the  sum-rule region of validity: $M^2=\{1.65, ~1.75,~1.9,~2.05,~2.2\}\,{\rm GeV^2}$.  The pole scheme has been used for the charm quark mass.}
\label{continuum_pole_fig}
\end{figure}

\begin{figure}[hbt]
\centering
\includegraphics[scale=0.8]{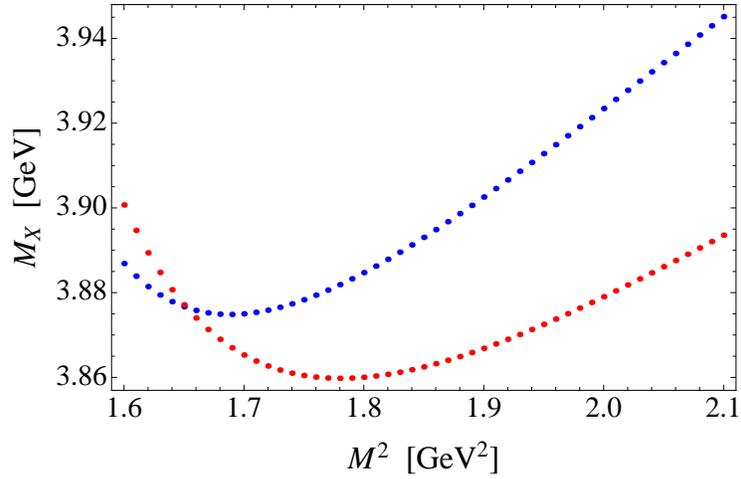}
\caption{ The sum-rule prediction of $M_X$ as a function of the Borel scale $M$ for $\xi=0.007$ and the optimized continuum $s_0=21\,{\rm GeV^2}$ (blue curve) and a slightly smaller value $s_0=19\,{\rm GeV^2}$ (red curve).
The range of $M$ corresponds to the sum-rule region of validity $1.6\,{\rm GeV}^2<M^2<2.1\,{\rm GeV}^2$ for the central $s_0$ values. The pole scheme has been used for the charm quark mass. }
\label{mass_pole_fig}
\end{figure}

We have chosen to use the ${\overline{\rm MS}}$ charm mass to align with the pure molecular and hybrid analyses \cite{Harnett:2012gs,Nielsen:2008}.  However, since these analyses and our mixing correlator calculation are leading order, there is no field-theoretical distinction between the pole and ${\overline{\rm MS}}$ scheme masses (see e.g., Ref.~\cite{Jamin:2001fw} for a next-to-leading order sum-rule analysis that employs both the pole and ${\overline{\rm MS}}$  schemes). Nevertheless, we explore the 
effect of using the pole-scheme charm mass $M_c=1.6\,{\rm GeV}$ \cite{Beringer:1900zz} as a source of theoretical uncertainty.  
In this case the mass in the pure molecular limit ($\xi=0$) is  significantly above the X(3872), and decreases with increasing $\xi$ until reaching 
 the optimized mixing parameter 
$\xi=0.007$ (see Figs.~\ref{continuum_pole_fig} and \ref{mass_pole_fig}).  There is minimal effect on the sum-rule window and the continuum.  Thus in the pole scheme, mixing effects  seem to be necessary for consistency with the X(3872).

\clearpage
\section{Conclusions}
\label{theConclusion}

In summary,
we have calculated the mixed correlation function of  $1^{++}$ hybrid charmonium and molecular currents, enabling a QCD sum-rule analysis of the X(3872) as a mixed state coupling to both
hybrid and molecular currents.
In the range of mixing parameters near the pure molecular limit, the largest mass prediction is in good agreement with X(3872) and represents a significant mixing of hybrid and molecular currents.
In general, the mixed current increases the stability of the sum-rule analysis compared to the pure molecular case \cite{Matheus:2006xi,Chen:2012pe,Chen:2010ze,Nielsen:2008}.  A mixed scenario where the X(3872) couples to a mixture of hybrid and molecular currents is thus viable.  However, as mentioned earlier, the sum-rule molecular and tetraquark currents are related by Fierz transformations and hence the interpretation of the underlying quark configuration is ambiguous. We anticipate that use of tetraquark currents would not lead to a substantial change in our conclusions because the molecular and tetraquark QCD sum-rule analysis mass predictions are  virtually indistinguishable \cite{Narison:2010pd}.

  Since only the four-quark state with $I=0$ can mix with the hybrid,  the mixing pattern may have a role in explaining  the X(3872) isospin-violating decay puzzle because the hybrid component 
  $D^* \bar D$ decay channel is highly suppressed \cite{page}
 and  will mainly decay into $J/\psi\omega$ \cite{Suzuki:2005ha}.  The standard explanation of isospin violation in X(3872) decay is that, for some non-perturbative reasons,   quarks like to be in 
  diquark (or meson) pairs, so that $[cu]$ and $[cd]$ diquarks (or meson clusters)  in the multi-quark system can be considered as a quasiparticle.  For instance, in \cite{Maiani:2004vq},
    $[cu][\bar c\bar u]$ and $[cd][\bar c\bar d]$  are independently bound. This pattern  predicts two neutral ``X(3872)" states with a  small mass difference, and charged partners of the X(3872)
    \cite{Maiani:2004vq,Faccini:2013lda}. 
 Similarly, the molecular scenario explains isospin violation in the decays \cite{Swanson:2003tb} and  predicts a second neutral state with a larger mass  mass splitting between the charged and neutral states  (see e.g., Ref.~\cite{Faccini:2013lda} for a recent discussion).
 Previously,  the charged partners of X(3872) had not been observed \cite{Choi:2011fc,exper}, but there is now evidence for their existence \cite{Ablikim:2013mio,Liu:2013dau,Xiao:2013iha}. Note that the experimental mass difference from the X(3872) is about 20~MeV, which does not seem to fit  either the tetraquark or molecular expectation.

 The discovery of the neutral partner of the X(3872) is thus crucial for  four-quark models of the X(3872).  In both models, the neutral partners are composed of $I=0,1$ states, so a hybrid mixture in the $I=0$ component provides a new mechanism for a mass splitting between the neutral states.
However, future work is needed to 
determine how a mixed charmonium hybrid and $\bar D D^*$  molecular scenario would 
affect the mass splitting of the neutral states
and whether it would
accomodate the observed isospin violation in  X(3872) decays \cite{Abe:2005ix,Choi:2011fc,Abe:2005iya}.

\bigskip
\noindent
{\bf Acknowledgements:}  TGS is grateful for financial support from the Natural Sciences and Engineering Research Council of Canada (NSERC). Hongying Jin is  is grateful for financial support from  NNSFC under grant 11175153/A050202.

\end{document}